\documentclass[aps,twocolumn]{revtex4-1}
\usepackage{graphicx}
\usepackage{subcaption}
\usepackage{amsmath}
\usepackage{amsfonts}
\usepackage{amsthm}
\usepackage{amssymb}
\usepackage{amsbsy}
\usepackage{wasysym}
\usepackage{bm}
\usepackage{mathrsfs}
\usepackage{color}
\usepackage{times}
\usepackage{hyperref}
\usepackage[resetlabels]{multibib}
\begin{document}
	
	\title{Crossover in growth law in the vapor-liquid phase separation inside complex porous medium}
	\author {Preethi M and Bhaskar Sen Gupta}
	\email{bhaskar.sengupta@vit.ac.in}
	\affiliation{Department of Physics, School of Advanced Sciences, Vellore Institute of Technology, Vellore, Tamil Nadu - 632014, India}

\begin{abstract}
We employ molecular dynamics simulations to investigate the domain morphology and growth kinetics of a vapor-liquid system embedded within a complex porous medium. By systematically varying the pore structure, we analyze the scaling behavior of correlation functions, structure factors, and domain growth exponents. The structure factor confirms the breakdown of Porod law and the emergence of fractal-like domain boundaries. Our key finding is the clear crossover in the domain growth law, from the classical power-law behavior observed in bulk fluids to a slower, logarithmic regime in highly confined systems. This transition is driven by energy barriers introduced by the porous geometry, which inhibit coarsening dynamics at later time. We provide a scaling analysis which further confirms this crossover and quantitatively connects the growth behavior with the average pore size.

\end{abstract}
	
\maketitle

\section{Introduction}

The phenomenon of vapor-liquid phase separation within porous media is ubiquitous in various natural and industrial processes~\cite{Gelb,Adidharma}. It plays a crucial role in enhanced oil recovery, geothermal energy production, geological carbon sequestration, and the performance of microfluidic devices, to name a few. In these applications, the intricate interplay between the phase transition and the fluid flow within the confined pore spaces significantly influences the overall efficiency and effectiveness of the process. Understanding the kinetics of this phase separation, i.e., the rate at which the vapor and liquid phases separate and evolve within the porous medium, is therefore of paramount importance. This knowledge is essential for accurate modeling, prediction, and optimization of these processes. However, there are fundamental questions that remain unanswered.

The kinetics of vapor-liquid phase separation in bulk systems has been extensively studied and well understood (see~\cite{Majumder}, and the references therein). Domain growth is characterized by a single time-dependent length scale $\ell(t)$, representing the average domain size that typically grows in a power law fashion $\ell(t) \sim t^\alpha$~\cite{Puri-book,Onuki,Binder-book,Jones}. The exponent $\alpha$ depends on the transport mechanism that drives phase separation. In the early stages, domain growth is diffusion-driven and $\alpha=1/3$ following the Lifshitz-Slyozov law~\cite{Lifshitz}. This is followed by a linear viscous hydrodynamic growth with $\alpha=1$~\cite{Majumder,Daniya-Gravity}. The above values of the growth exponent are universal and pertinent to bulk systems.

 Challenges arise when confinement is imposed, stemming from the intricate interplay between phase separation and geometric constraints~\cite{Gelb}. The presence of pore walls, constrictions, and varying pore sizes in porous media significantly modifies these processes. Capillary forces, surface tension, and wetting characteristics become dominant factors that influence nucleation sites, interfacial curvature, and mass transport. The topology of the porous network, including connectivity and tortuosity, further complicates the dynamics of phase separation by restricting domain growth and altering the pathways for mass transport and effective diffusivity~\cite{Adidharma}. These factors can lead to deviations from the classical scaling laws observed in bulk systems.

 In this paper, we focus on elucidating the kinetics of vapor-liquid phase separation and domain growth within complex porous media for a one-component fluid system using computer simulations. We specifically address how the liquid and vapor domains evolve, the scaling laws governing their growth with time, and the influence of different pore structures on these dynamics compared to bulk systems, which to the best of our knowledge have remained unexplored.

\section{Model and method}
 While continuum-scale models offer valuable insight~\cite{Ngamsaad}, they often struggle to capture the complex interplay of molecular interactions and interfacial phenomena that dominate at the pore scale. Molecular dynamics (MD) simulations, with their atomistic resolution, provide a powerful tool for investigating these processes at the molecular level. In this study, we resort to MD simulations using LAMMPS software~\cite{lammps} to study the kinetics of vapor-liquid phase separation imbibed in complex porous media.  

To create the porous medium, we consider a binary mixture system made up of two distinct particle species, A and B, present in equal quantities (50:50). The system is set up at a high particle density $\rho=N/V=1$, where $N$ is the number of particles and $V$ is the volume of the cubic system. Particle interactions are controlled by the standard Lennard-Jones (LJ) potential, expressed as:
\begin{equation}
    U_{\alpha\beta}(r) = 4\epsilon_{\alpha\beta} \left[\left(\frac{\sigma_{\alpha\beta}}{r}\right)^{12} - \left(\frac{\sigma_{\alpha\beta}}{r}\right)^6\right]
    \label{1}
\end{equation}
where $\sigma_{\alpha\beta}$ and $\epsilon_{\alpha\beta}$ represent the particle diameter and interaction strength, respectively. The indices $\alpha, \beta \in \{A, B\}$ and $r = |\vec{r}_i - \vec{r}_j|$ represents the separation distance between two particles. To improve computational efficiency, the interaction range is restricted to $r_c$ = 2.5$\sigma$. The interaction parameters are set to $\epsilon_{AA} = \epsilon_{BB} = 2\epsilon_{AB} = 1.0$ and $\sigma_{AA} = \sigma_{BB} = \sigma_{AB} = 1.0$ to ensure phase separation between the two species. These values correspond to a critical temperature $T_c$ = 1.42, which ensures liquid-liquid phase separation and the system avoids liquid-solid or gas-liquid transitions~\cite{criticalT}. The length and temperature are measured in units of $\sigma$ and $\epsilon/k_B$, respectively, $k_B$ being the Boltzmann constant. For convenience, the mass of each particle and $k_B$ are set to unity. The periodic boundary condition is applied in all three directions.

Our simulation begins by equilibrating the system of $N$ = 262,144 particles at a high temperature T = 10.0 using MD simulations to prepare a homogeneous mixture. During MD simulations, the position and velocity of the particles are updated using the velocity-Verlet integration technique~\cite{verlet} with a time step $\Delta t = 0.005$. Here, time is measured in units of $\sqrt{{m\sigma^2/\epsilon}}$. Subsequently, the system is quenched at a lower temperature T = 0.77$T_c$ at time $t=0$. Following that, the binary mixture starts to phase separate, creating bicontinuous A-rich and B-rich domains. After a time period $\tau$, the domains formed by any type of species (say $A$-type) are designated as porous media. The $A$-type particles are kept frozen in their respective positions throughout the rest of the simulation, and the A species acts as a host structure. 

The other species ($B$-type) is manipulated to create the fluid phase. Appropriate number of randomly chosen $B$-type particles is then removed from the simulation box to achieve a target density $\rho = 0.3$ of the same. The LJ interaction between cross species is truncated at $r_c = 2^{1/6}\sigma$, which eliminates the attractive forces between the frozen and fluid particles. This setup ensures a well-defined fluid system within the porous host. The described fluid system has a bulk critical temperature of $T_c = 0.94\epsilon/k_B$ and a critical density $\rho_c = 0.32$ associated with the vapor-liquid transition~\cite{criticalT-vapor}. To annihilate any residual memory, the fluid is reheated to T = 10.0 and equilibrated. Finally, the system is quenched to a temperature T = 0.8. Throughout the simulation, the Nosé-Hoover thermostat is used to control the temperature while preserving hydrodynamic effects~\cite{Nose}. This setup enables the study of phase separation, domain growth, and coarsening dynamics of the vapor-liquid transition within a confined porous environment. The ensemble average of all the statistical quantities is obtained from 30 independent runs starting from completely different initial configurations.

To study the dynamics of phase ordering, we resort to the length scale $\ell(t)$ representing the average size of the domains. This is computed from the two-point equal-time correlation function~\cite{Bray}, given by
\begin{equation}
C(\vec{r}, t) = \langle \psi(0, t)\psi(\vec{r}, t) \rangle - \langle \psi(0, t) \rangle \langle \psi(\vec{r}, t) \rangle
\label{2}
\end{equation}
where $\langle . \rangle $ denotes the statistical averaging, and $ \psi(\vec{r}, t)$ represents the order parameter. By mapping the system onto an Ising model, the order parameter $ \psi(\vec{r}, t)$ is expressed as:
\begin{equation}
    \psi(\vec{r}, t) =
\begin{cases} 
 0 & \text{frozen particles},\\
+1 & \text{if } \rho(\vec{r}, t) > \rho_c, \\
-1 & \text{otherwise},
\end{cases}
\label{3}
\end{equation}
where $\rho(\vec{r}, t)$ is the local density over a box of size $(2\sigma)^3$ at position $\vec{r}$~\cite{Bhattacharyya1,Davis1,Parameshwaran1}.

The interface morphology can be analyzed using the structure factor  $S(\vec{k}, t)$, which is obtained from the Fourier transform of the correlation function~\cite{Bray,Parameshwaran2}:
\begin{equation}
    S(\vec{k}, t) = \int d\vec{r} \, \exp(i \vec{k} \cdot \vec{r}) \, C(\vec{r}, t)
    \label{4}    
\end{equation}
Finally for the isotropic system, spherically averaged $C(r,t )$ and $S(k,t )$ are computed.

\section{Results}
\begin{figure}[ht]
  \centering
     \includegraphics[width=26mm]{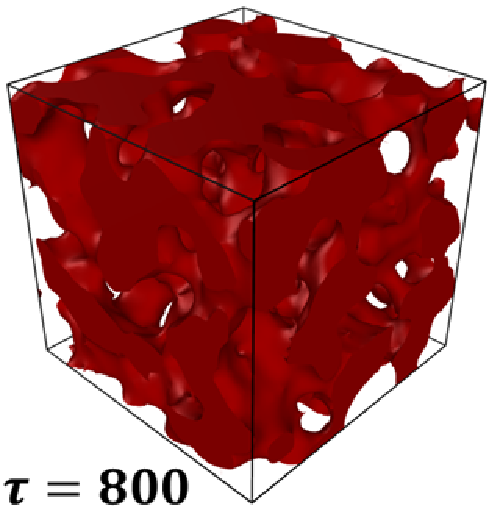}
      \includegraphics[width=26mm]{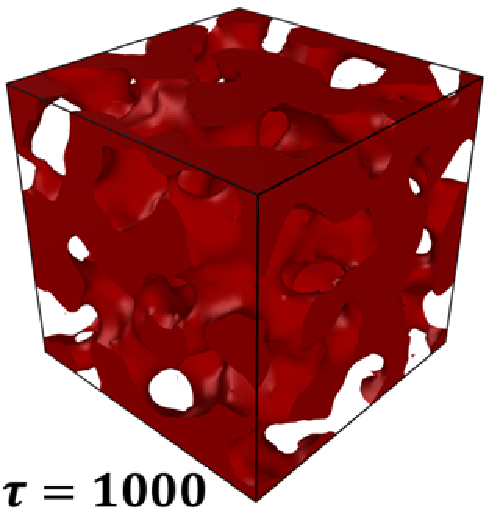}
       \includegraphics[width=26mm]{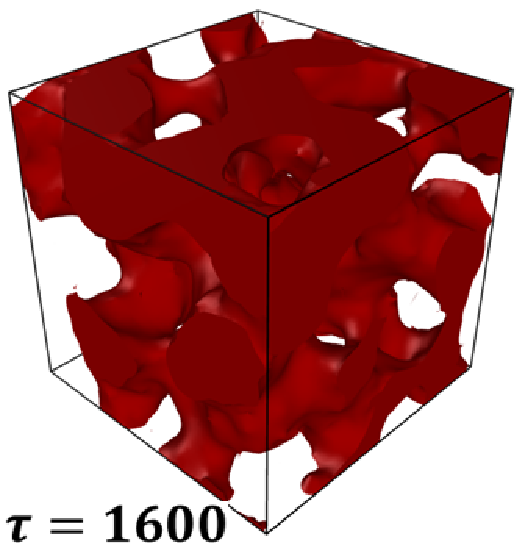}\\
        \includegraphics[width=26mm]{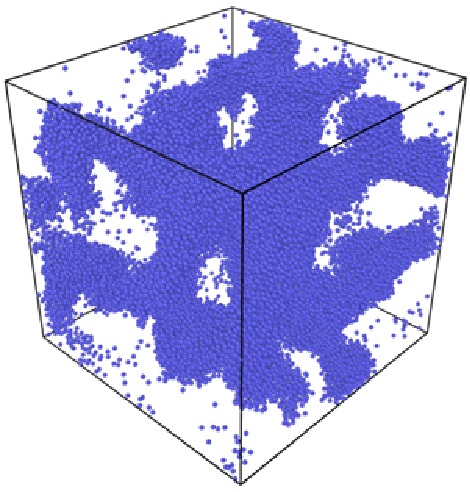}
         \includegraphics[width=26mm]{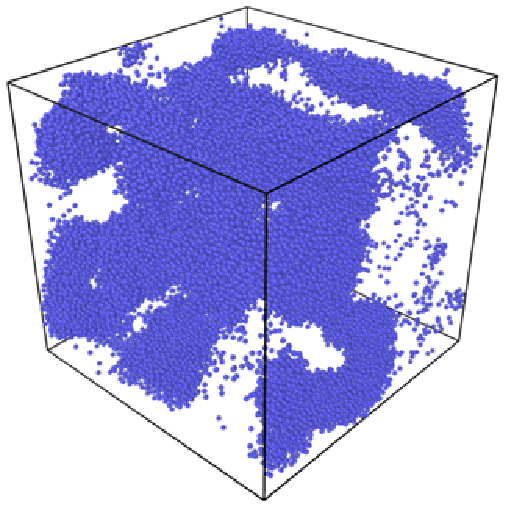}
          \includegraphics[width=26mm]{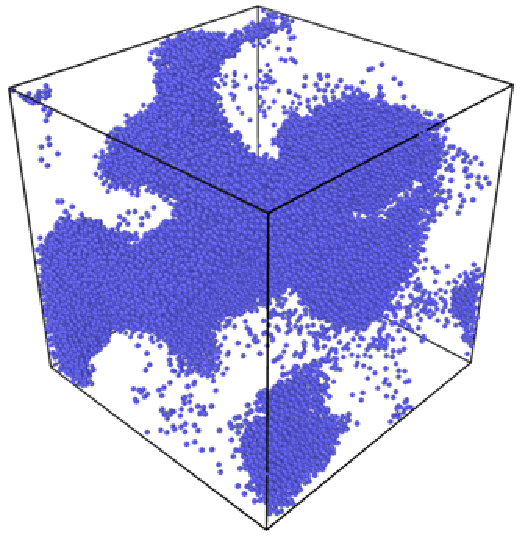}
     \caption{Upper panel: Three different porous host structures corresponding to different $\tau$ values mentioned in the figure. Lower panel: Typical snapshots of the phase-separating vapor-liquid systems at time $t$ = 500.}
     \label{fig1:snap}
 \end{figure}

The simulation protocol dictates a direct relationship between the time period $\tau$ and the average pore size ($d_P$), and the $d_P$ increases with $\tau$. We consider three distinct porous host structures that correspond to $\tau = 800, 1000,$ and $1600$, as shown in Fig.~\ref{fig1:snap}. 
It is conspicuous that the average pore size increases with $\tau$. Fig.~\ref{fig1:snap} illustrates the effect of these porous media on phase separation dynamics. The snapshots reveal that, at a given time $t=500$, the cluster sizes of the liquid phase increase with $d_P$. Therefore, the phase separation kinetics is strongly dependent on $d_P$.

\begin{figure}[ht]
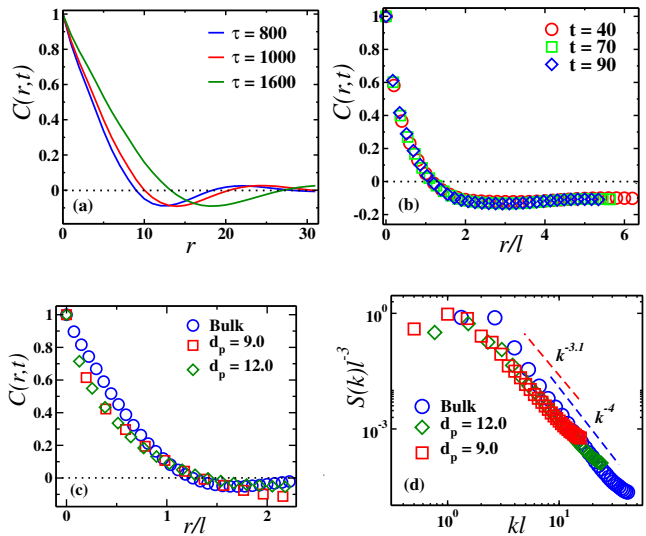

	\centering
	\includegraphics[width=41mm]{Fig2-Poresize.eps}~
	\includegraphics[width=41mm]{Fig3-ScaledCr.eps}\\
    \vspace{.5cm}
	\includegraphics[width=42mm]{Fig4-scaledCr_diffPore.eps}~
	\includegraphics[width=41mm]{Fig5-Sk.eps}	
	\caption{ (a) Correlation function $C(r,t)$ vs. $r$ for the frozen $A$-type particles constituting three diﬀerent porous host structures. (b) The scaling plot of $C(r,t)$ vs. $r/\ell$ for the fluid system confined within the porous material with $d_p$ = 9.0 at different time. (c) The same scaling plot in (b) for the bulk and confined fluid systems at time $t=100$. (d) The scaled structure factor $S(k)\ell^{-3}$ vs. $k\ell$ plot for the systems in (c) at time $t$ = 100.}
	\label{fig:2}
\end{figure}

To qualitatively analyze the average domain size, we compute the two-point equal-time correlation function $C(r,t)$ given by Eq. (\ref{2}). We first compute the average pore size of the host materials shown in Fig.~\ref{fig1:snap}, which corresponds to the average domain size of the frozen $A$-type particles. Fig.~\ref{fig:2}(a) illustrates $C(r,t)$ for the $A$-type particles as a function of distance $r$ for $\tau = 800, 1000,$ and $1600$. Throughout the paper, we utilize the first zero cross of $C(r,t)$ as a reliable measure of the average domain size $\ell(t)$~\cite{Bray}. We find that the average pore sizes $d_p$ ($=\ell(t)$ in this case) are approximately 8.0, 9.0 and 12.0 for $\tau = 800$, 1000 and 1600 respectively.

In Fig.~\ref{fig:2}(b), we show the scaling behavior of the correlation function $C(r,t)$ plotted against $r/\ell(t)$ for a chosen pore size of $d_p$ = 9.0. A neat data collapse for different times demonstrates the self-similar nature of domain growth~\cite{Binder-book}. A similar behavior is also observed for other porous structures corresponding to $d_p$ = 8.0 and 12.0 (not shown here). This consistency suggests that the phase-separating fluid, even when confined within a porous medium, falls within the same dynamical universality class, meaning that its scaling properties remain vindicated.

In Fig.~\ref{fig:2}(c), we show the scaling plots of $C(r,t)$ vs. $r/\ell(t)$ corresponding to the bulk and confined systems at a fixed time. The data collapse is clearly inadequate, indicating that the scaling does not hold in this case. Therefore, the super-universality property does not apply to the confined system~\cite{CorberiSU1,CorberiSU2}. For the bulk system, the function exhibits linear decay with $r$ over small distances, a characteristic attributed to scattering from sharp interfaces, commonly referred to as the Porod law~\cite{Bray}. In contrast, confined systems display a non-linear cusp in the short distance limit, indicating a fractal interface and the violation of the Porod law~\cite{Gaurav,Bhattacharyya2,Bhattacharyya3}. 

To gain further insight into the morphology of the domain boundaries, we focus on the static structure factor $S(k,t)$. In Fig. \ref{fig:2}(d), we show the scaled $S(k,t)$ plots associated with the correlation functions in Fig. \ref{fig:2}(c) for the bulk and confined systems. The pore size-dependent scaling confirms the breakdown of superuniversality~\cite{CorberiSU1,CorberiSU2}. Although the bulk system follows Porod’s law, characterized by $S(k,t) \sim k^{-(d+1)}$ in the large $k$ limit~\cite{Puri-book}, the effect of confinement yields a clear departure from this behavior, indicating a non-Porod tail, $S(k,t) \sim k^{-(d+\theta)}$ where $\theta$ is approximately equal to 0.1 for $d_P=9.0$. This reconfirms the roughening of the domain interfaces with fractal dimension $d_f= d-\theta=2.9$.

\begin{figure}[t]
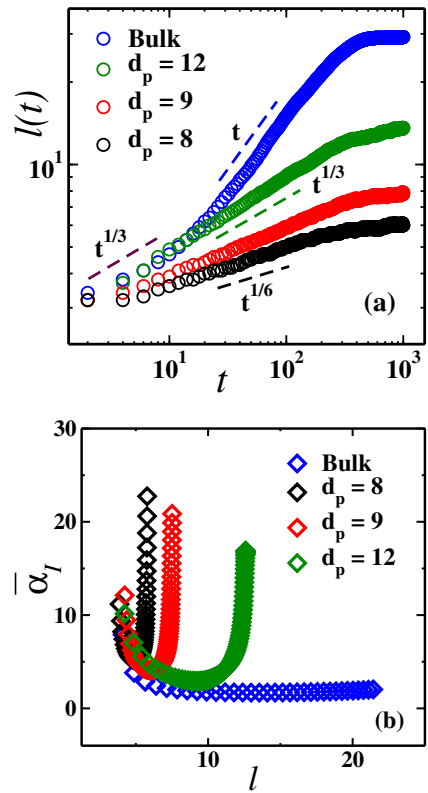

  \centering
  \includegraphics[width=55mm]{Fig6-lengthscale.eps}\\
  \vspace{0.3cm}
  \includegraphics[width=55mm]{Fig7-zeff.eps}
     \caption{The time evolution of the average domain size $\ell(t)$  for the bulk and confined system, depicted on a log-log scale. Inverse of the instantaneous dynamical exponent $\bar{\alpha_I}$ vs. $\ell$ for the length scale shown in (a).}
     \label{fig:3}
 \end{figure}

Next, we quantify the domain growth dynamics in terms of the length scale $\ell(t)$ and explore how it varies between bulk and confined systems. Fig.~\ref{fig:3}(a) presents the time dependence of the average domain size $\ell(t)$. Our simulations successfully capture the linear viscous hydrodynamic growth $\ell(t) \sim t$ after the transient diffusive regime $\ell(t) \sim t^{1/3}$ for the bulk system. 
 
A dramatic slowing down in the growth dynamics is observed for systems confined within porous media due to the influence of complex pore structures, as observed in Fig.~\ref{fig:3}(a). This slowing is more pronounced as the pore size decreases, indicating that the confinement disrupts the natural coarsening process. The long-time behavior of the length scale indicates that complete phase separation is not possible and the system reaches a metastable state as the average pore size decreases. This is because the presence of confinement creates additional barriers to the movement of the initially formed disconnected vapor and liquid regions and hinders their coalescence. Consequently, a modified domain growth law is anticipated for the confined system. 

\begin{figure}[t]
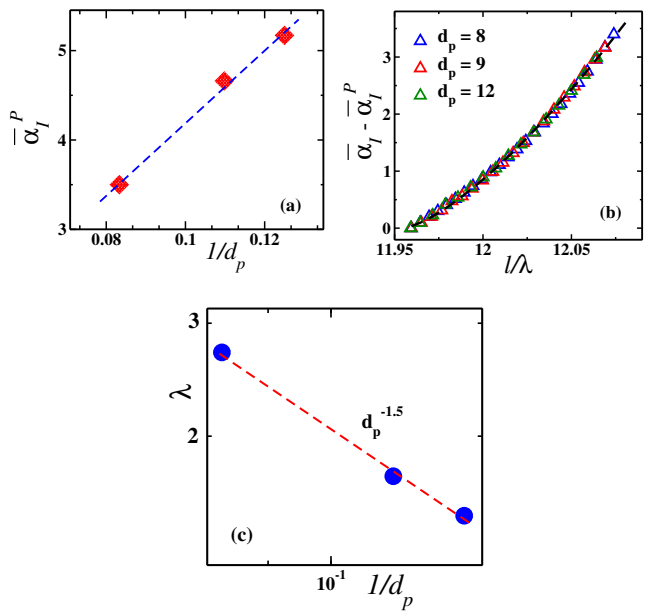

	\centering
	\includegraphics[width=42mm]{Fig8-z_bar.eps}~~
	\includegraphics[width=40mm]{Fig9-MasterCurve.eps}\\
    \vspace{0.4cm}
    \includegraphics[width=42mm]{Fig10-slope.eps}	
	\caption{(a) The variation of $\bar{\alpha}^P_I$ with $1/d_p$ (see text). Dashed line represents the best linear fit. (b) ($\bar{\alpha}_I - \bar{\alpha}^P_I$) is plotted as a function of $\ell/\lambda$ for different $d_p$ values (see text). The dashed line represents the best fit curve following Eq.~\ref{scaling}. (c) The variation of $\lambda$ with $1/d_p$ in the log-log scale. Dashed line represents the best fit power law curve with exponent -1.5.}
	\label{fig:4} 
\end{figure}

To examine for any modification of the growth law influenced by the porous medium, we compute the instantaneous slope of the length scale given as $\alpha_I = \frac{d\ln(\ell)}{d\ln(t)}$. In Fig.~\ref{fig:3}(b) we show the variation of $\bar{\alpha_I}=1/\alpha_I$ with $\ell(t)$. As expected, $\bar{\alpha_I}$ saturates to unity in the long time limit for the bulk system, which corresponds to the viscous hydrodynamic regime. In contrast to this, confined systems exhibit a flat line in the graph $\bar{\alpha_I}$ vs. $\ell(t)$ for a brief intermediate time, where the growth obeys a power-law behavior. The value of $\bar{\alpha_I}$ (say, $\bar{\alpha}_I^P$) corresponding to this intermediate power law regime is shown with $1/d_P$ in Fig.~\ref{fig:4}(a) (note that the barrier increases with $1/d_P$). From the figure, we observe a linear relation between these two quantities. Therefore, the energy barrier created by the complex porous structures that are responsible for slowing down the phase separation dynamics has a logarithmic dependence on the domain size $\ell(t)$~\cite{Paul1,Paul2}.
 
From Fig.~\ref{fig:3}(a) it is evident that at the late time there is a crossover regime where the length scale saturates asymptotically and the growth rate comes to a halt. The saturation length scale strongly depends on the degree of confinement. We analyze this crossover regime, following the method described in~\cite{CorberiSU1,CorberiSU2}.  Therefore, we plot $\bar{\alpha_I}-\bar{\alpha}_I^P$ as a function of $\ell/\lambda$ in Fig.~\ref{fig:4}(b), where $\lambda$ is a parameter dependent on $d_p$. The parameter $\lambda$ is chosen to ensure a neat collapsing of the data between different pore sizes. In Fig.~\ref{fig:4}(c) we show the variation of $\lambda$ with $1/d_P$. The fitting parameter is found to depend on the pore size as $\lambda \sim d_p^{-1.5}$. The negative exponent refers to a crossover to a logarithmic behavior~\cite{CorberiSU1,CorberiSU2}. Therefore, the constraint imposed by the porous medium alters the coarsening dynamics, and the corresponding growth law shows a clear crossover from power law to logarithmic behavior. 

We consider the following power-law function to explain the scaling behavior of the data shown in Fig. \ref{fig:4}(b)~\cite{Ahmed}
\begin{equation}\label{scaling}
\bar{\alpha_I} = \frac{d~\mathrm{ln} (t)}{d~\mathrm{ln} (\ell)} = \bar{\alpha}_I^P + a {(\frac{\ell-\ell_0}{\lambda})^\phi}
\end{equation}
where $a \simeq 2.6, \ell_0 \simeq 4.969$ and $\phi \simeq 1.6$ (logarithmic growth exponent) are the fitting parameters. Eq.~\ref{scaling} yields the logarithmic growth law $\frac{\ell}{\lambda} \sim (\frac{\phi}{b} \mathrm{ln} t)^{1/\phi}$. The scaling function fits well with the numerical data (dashed line in Fig. \ref{fig:4}(b)), providing analytical validation of the crossover behavior. These findings demonstrate the critical role of pore size in determining the system's growth dynamics, particularly during the transition from power law to logarithmic scaling. A similar crossover behavior was observed in the liquid-liquid phase separation of binary fluids in the presence of a quenched random field~\cite{Ahmed}. Our observation is also consistent with the analytical study of phase separation kinetics in fluids confined within a porous medium, which employed coupled partial differential equations, specifically the Cahn-Hilliard and Brinkman-Darcy equations~\cite{Ngamsaad}.

\section{Conclusions}
In this study, we performed extensive molecular dynamics simulations to investigate vapor-liquid phase separation kinetics within complex porous media. By systematically varying the pore size of the confining structure, we demonstrated a strong dependence of domain growth on the underlying geometry of the host medium. Analysis of correlation functions and structure factors reveals that the presence of confinement alters the interfacial morphology, evidenced by the breakdown of Porod’s law, suggesting the roughening of domain interfaces. Our results revealed a pronounced slowdown in phase separation as a result of confinement. The growth exponent exhibited a direct dependence on the pore size leading to a crossover from the classical power-law growth observed in bulk systems to a logarithmic regime at later times, driven by the energy barriers imposed by the porous structure. An appropriate scaling analysis was presented to justify the given observations. These results demonstrated the significant impact of geometrical constraints on phase-separation kinetics. It will be worth extending the current framework to explore additional aspects by incorporating the wetting interaction. Finally, it will be interesting to study the scaling laws of domain coarsening and the aging dynamics under diﬀerent wetting conditions.

\textit{Acknowledgements.}--B. Sen Gupta acknowledges the Science and Engineering Research Board (SERB), Department of Science and Technology (DST), Government of India (no. CRG/2022/009343) for financial support. Preethi M. acknowledges DST-INSPIRE, India for doctoral fellowship.


\begin{thebibliography}{99}
\bibitem{Gelb}
L. D. Gelb, K. E. Gubbins, R. Radhakrishnan and M. Sliwinska-Bartkowiak, Rep. Prog. Phys. \textbf{62}, 1573 (1999).
\bibitem{Adidharma}
H. Adidharma and S. P. Tan, Ind. Eng. Chem. Res. \textbf{61},15488 (2022). 	
 \bibitem{Majumder}
 S. Majumder and S. K. Das, Europhys. Lett. \textbf{95} 46002 (2011).
\bibitem{Puri-book} 
S. Puri and V. Wadhawan, Kinetics of Phase Transitions (CRC Press, Boca Raton, FL, 2009).
\bibitem{Onuki} 
A. Onuki, Phase Transition Dynamics, Cambridge University Press, Cambridge, (2002).
\bibitem{Binder-book} 
K. Binder and P. Fratzl, Phase Transformation in Materials, ed. G. Kostorz, Wiley, Weinheim, p. 409 (2001).
\bibitem{Jones}
R. A. L. Jones, Soft Condensed Matter, Oxford University Press, Oxford, (2008).	
\bibitem{Lifshitz}
I. M. Lifshitz and V. V. Slyozov, J. Phys. Chem. Solids \textbf{19}, 35 (1961).
\bibitem{Daniya-Gravity}
D. Davis and B. Sen Gupta, Soft Matter \textbf{21}, 1012 (2025).
\bibitem{Ngamsaad}	
W. Ngamsaad, J. Yojina and W. Triampo, J. Phys. A: Math. Theor. \textbf{43}, 202001 (2010).
 \bibitem{lammps}
 A. P. Thompson, H. M. Aktulga, R. Berger,  D. S. Bolintineanu, W. M. Brown, P. S. Crozier, P. J. in 't Veld, A. Kohlmeyer, S. G. Moore, T. D. Nguyen, R. Shan, M. J. Stevens, J. Tranchida, C. Trott and S. J. Plimpton, Comp. Phys. Comm. \textbf{271}, 108171 (2022).
 \bibitem{criticalT}
 S. K. Das, M. E. Fisher, J. V. Sengers, J. Horbach, and K. Binder, Phys. Rev. Lett. \textbf{97}, 025702 (2006).
 \bibitem{verlet}
 L. Verlet, Phys. Rev. \textbf{159}, 98 (1967).
\bibitem{criticalT-vapor}
S. Roy and S. K. Das, Phys. Rev. E \textbf{85}, 050602 (2012).
\bibitem{Nose}
D. Frenkel and B. Smit, Understanding Molecular Simulations: From Algorithms to Applications (Academic Press, San Diego, 2002).
\bibitem{Bray} 	
A. J. Bray, Adv. Phys., 51, 481 (2002). 	
\bibitem{Bhattacharyya1}
R. Bhattacharyya and B. S. Gupta, Europhys. Lett. \textbf{140}, 47002 (2022).
\bibitem{Davis1}
D. Davis and B. Sen Gupta, Phys. Rev. E \textbf{108}, 064607 (2023).
\bibitem{Parameshwaran1}
A. Parameshwaran, and B. Sen Gupta, Phys. Rev. E \textbf{111}, 025405 (2025).
\bibitem{Parameshwaran2}
D. Davis, A. Parameshwaran, and B. Sen Gupta, arXiv:2412.02774.
\bibitem{CorberiSU1}
F. Corberi, E. Lippiello, A. Mukherjee, S. Puri, and M. Zannetti, J. Stat. Mech. P03016 (2011).
\bibitem{CorberiSU2}
F. Corberi, E. Lippiello, A. Mukherjee, S. Puri, and M. Zannetti, Phys. Rev. E \textbf{85}, 021141 (2012).
\bibitem{Gaurav}
G. P. Shrivastav, S. Krishnamoorthy, V. Banerjee, and S. Puri, Europhys Lett. \textbf{96}, 36003 (2011)
\bibitem{Bhattacharyya2}
R. Bhattacharyya and B. S. Gupta, Phys. Rev. E \textbf{104}, 054612 (2021).
\bibitem{Bhattacharyya3}
R. Bhattacharyya and B. S. Gupta, Soft Matter \textbf{20}, 2969 (2024).
\bibitem{Paul1}
R. Paul, S. Puri, and H. Rieger, Europhys. Lett. \textbf{68}, 881 (2004).
\bibitem{Paul2}
R. Paul, S. Puri, and H. Rieger, Phys. Rev. E \textbf{71}, 061109 (2005).
\bibitem{Ahmed}
S Ahmad, S. Puri, and S. K. Das, Phys. Rev. E \textbf{90}, 040302(R) (2014).

	
\end{thebibliography}
\end{document}